\documentclass[runningheads]{llncs}

\usepackage[utf8]{inputenc}
\usepackage[numbers, comma, sort]{natbib}
\usepackage{amssymb}
\usepackage{graphicx}
\usepackage{float}

\usepackage{wrapfig}

\usepackage{pgfplots}
\usepackage{tikz}
\usetikzlibrary{arrows,backgrounds,positioning}

\usepackage{url}
\urldef{\mailsa}\path|rob@clabs.cc,schubotz@tu-berlin.de |
\newcommand{\todaydate}{\leadingzero{\day}.\leadingzero{\month}.\the\year}
\newcommand{\keywords}[1]{\par\addvspace\baselineskip
\noindent\keywordname\enspace\ignorespaces#1}

\begin{document}

\mainmatter

\title{Mathematical Language Processing \\ Project}
\titlerunning{MLP}

\author{Robert Pagel \and Moritz Schubotz}
\authorrunning{Pagel, Schubotz}

\institute{
Database Systems and Information Management Group,\\
Technische Universit\"{a}t Berlin,
Einsteinufer 17, 10587 Berlin, Germany\\
\mailsa\\
\url{http://demo.formulasearchengine.com/}}

\maketitle

\begin{abstract}

In natural language, words and phrases themselves imply the semantics. In
contrast, the meaning of identifiers in mathematical formulae is undefined.
Thus scientists must study the context to decode the meaning. The Mathematical
Language Processing (MLP) project aims to support that process. In this
paper, we compare two approaches to discover identifier-definition tuples. At
first we use a simple pattern matching approach. Second, we present the MLP
approach that uses part-of-speech tag based distances as well as sentence
positions to calculate identifier-definition probabilities. The evaluation of
our prototypical system, applied on the Wikipedia text corpus, shows that our
approach augments the user experience substantially. While hovering the
identifiers in the formula, tool-tips with the most probable definitions
occur. Tests with random samples show that the displayed definitions provide a
good match with the actual meaning of the identifiers.

\end{abstract}

\section{Introduction}

\begin{wrapfigure}{r}{0.4\textwidth}
\label{fig:screenshot}
\vspace{-20pt}
	\includegraphics[width=0.4\textwidth]{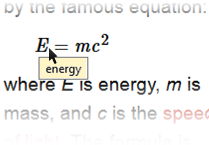}
\caption{Screenshot of the energy mass relation page `Mass–energy equivalence', while hovering the letter `E'.}
\vspace{-20pt}
\end{wrapfigure}

Mathematical formulae are viable sources of information for a wide range of
scientists. Often, they contain identifiers whose meaning might be at first
unknown or at least ambiguous to the reader (depending on their knowledge).
Therefore, one usually needs to study the surrounding text to find the
relevant definition. An automatic information retrieval system can be used to
reduce the reader's effort by displaying the most relevant definition relation
found to the reader. Students and scientists of other disciplines would
especially profit from a system that helps them to understand formulae more
quickly. In the long term, the extracted identifier definition tuples contribute
to an increased machine readability of scientific publications. This builds a
foundation for added value services such as search, clustering and improved
accessibility.

To build such a system, a labelled text corpus that annotates identifiers and
their definition is desirable. At the project start, such a corpus was not
available. Consequently we had to start manual investigation of individual
articles. Our first observation was that many identifier definitions use a
fixed string pattern to explain the definition to the reader. Furthermore,
most definitions usually appear very close to the related identifier in the
sentences. Thus, we calculate the probabilities for correct identifier
definition tuples based on distance metrics for certain part-of-speech (POS)
tagged words. This correlates to the experience that readers usually extract
identifier definitions from context that is given by the surrounding text.

We chose the Wikipedia as the target text corpus because of two facts. First,
most articles make use of \texttt{<math/>} tags (\emph{texvc} as an input
language) for formulae. The identification of \texttt{<math/>} tags is
trivial, and from the MathML output, it is easy to extract the identifiers.
Second, the articles are already annotated with mark-up. Particularly, hyperlinks
to other articles within Wikipedia are of interest as they typically wrap
around any number of words and indicate that these in combination are relevant
in the given context or (respectively) sentence.

The English Wikipedia contains roughly four million articles. Even if we only
pick articles containing \texttt{<math/>} tags, our processor still needs to
compute with tens of thousands of articles. Especially when using text
annotators (e.g., \emph{POS} tagger \cite{Rathna96}), like Stanford's NLP
framework, one can make use of a parallel processing system to speed up
computation. We implement the proposed strategy with the Stratosphere system
\cite{stratosphere}. It is based on the PACT programming model
\cite{Alexandrov2010}, which enables us to rapidly generate a large amount of
definition relation candidates with only minimal implementation overhead for
the parallelization.

\paragraph{Related Work.}

\citeauthor{Quoc2010} \cite{Quoc2010} proposed an approach for
relating whole formulas to sentences and their describing paragraphs.
\citeauthor{Yokoi} \cite{Yokoi} trained a \emph{support vector machine} to extract
natural language descriptions for mathematical expressions. Furhter work in this field was
done by \cite{ref2} and \cite{ref1}.

\section{Pattern-based Definition Discovery}

At first, we implemented a simple identifier definition extractor that is
based on a set of static patterns. As this is a fairly robust approach and easy
to implement, it serves as a good reference point in terms of
performance. It simply iterates through the text, trying to find word groups,
that are matched by a pattern. The patterns being used to discover description
terms are depicted in Table \ref{tpatterns}. Due to the fact that we already
tokenized and annotated the articles in a previous step in the MLP system, we
can make use of POS tags here as well.

Note, determiners not only contain articles, but also quantifiers and
distributives. The last pattern in Table \ref{tpatterns} contains `*/DT'. This
is a shorthand for every word, that has the POS tag `DT' (determiner).
Otherwise this pattern would be rather large, as it needs to contain every
possible determiner. \emph{IDENTIFIER} as well as \emph{DESCRIPTION} are
place-holder, that mark the positions of the entities from a possible definition
relation.

\begin{table}
\vspace{-5pt}
	\begin{center}
		\begin{tabular}{| p{9.3cm} |}
			\hline
			Pattern \\
			\hline
			\texttt{\emph{<description>} \emph{<identifier>}} \\
			\texttt{\emph{<identifier>} is \emph{<description>}} \\
			\texttt{\emph{<identifier>} is the \emph{<description>}} \\
			\texttt{let \emph{<identifier>} be the \emph{<description>}} \\
			\texttt{\emph{<description>} is|are denoted by \emph{<identifier>}} \\
			\texttt{\emph{<identifier>} denotes */DT \emph{<description>}} \\
			\hline
		\end{tabular}
	\end{center}
\caption{\label{tpatterns}Implemented static patterns}
\vspace{-5pt}
\end{table}

\section{Statistical Definition Discovery}

We detect relations between identifiers and their description in two steps.
First, we extract the identifiers from the formulae found in an article, and
second we determine their description from the surrounding text.

Extracting relevant identifiers from the article relies on the assumption that
the author will use \texttt{<math/>} tags for all formulae. This said, a
formula that is written in the running text cannot be recognized, and therefore,
cannot be extracted by our system.

The fact that we can estimate all relevant identifiers for an article (see
Section \ref{ir}), combined with some common assumptions about definition relations,
can be exploited to largely reduce the set of candidates that need to be
ranked. Please note that this reduction is essential for retrieving the
correct relations for our approach. Otherwise almost any word would be ranked
and the precision of the retrieval would drop significantly.

The basic assumption of our approach is that the two entities of a definition
relation co-occur in the same sentence. In other words, if we want to retrieve
the description for an identifier, only sentences containing the identifier
could include the definition relation. Having said this, any other sentences can
be ignored. Furthermore, we assume that it is more likely to find the
description in first sentences than in the latter. This is based on the idea
that authors introduce the meaning of an identifier and than subsequent use the
identifier, without necessarily repeating its definition.

Another assumption can be made about the lexical class of the definition relation we
want to rank. The descriptions are nouns or even noun phrases (e.g., \texttt{`the
effective spring constant k'} or \texttt{`mass m of something'}). We discard
all other words (according to their POS tag) except noun phrases and Wikipedia
hyper-links. These are the candidates descriptions for a definition relation. Noun phrases
and hyper-links may consist of multiple words. For all intents and purposes, it
is not necessary to threat noun phrases and hyper-links as a set of words, and
therefore, they will be treated subsequently as if they were one. This is
important, due to the fact that the overall ranking will be greatly influenced
by the distance of candidates to the position of the identifier.

\subsection{Numerical Statistics}

Each description candidate is ranked with the weighted sum

\begin{equation} \label{eq:rating}
R(n,\Delta,t,d)=\frac{\alpha{R}_{\sigma_\mathrm d}(\Delta)
+\beta{R}_{\sigma_\mathrm s}(n)
+\gamma\,\mathrm{tf}(t,s)}{\alpha+\beta+\gamma} \mapsto [0,1].
\end{equation}

The weighted sum depends on the distance $\Delta$ (amount of word tokens) between
identifier and the description term $t$, the sentence number $n$ counting
(from the beginning of the article) all sentences containing the identifier,
and the term frequency $\mathrm{tf}(t,s)$ in the set of sentences $s$. The
distance was normalized with $R_\sigma(\Delta) = \exp\left[ -\frac{1}{2}
\frac{\Delta^2-1}{\sigma^2}\right].$ We assume that the probability to find a
relation at $\Delta=1$ is maximal. For example in the text fragment
\texttt{`the energy E, and the mass m'}, in order to determine the full width
at half maximum of our distribution, we evaluated some articles manually and
found $R_{\sigma_\mathrm d}(1)\approx 2 R_{\sigma_\mathrm d}(5)$ and thus
$\sigma_d=\sqrt\frac{12}{\ln 2}$. The probability to find a correct definition
decays to 50\% within three sentences. Consequently $\sigma_\mathrm
s=2\left({\ln 2}\right)^{-\frac{1}{2}}$.

\paragraph{Robustness.}

The classic tf-idf \cite{Salton86} statistic reflects the importance of a term
to a document. For our task, the inverse document frequency (idf) assigns
high penalties to frequent words like `length', as opposed to words seldom
seen such as `Hamiltonian'. These are both valid definitions for identifiers.
As the influence of $\mathrm{tf}(t,s)$ on the sensitivity of the overall ranking
(\ref{eq:rating}) seems to be very high, we reduce the impact with the tuning
parameters $\gamma=0.1$ and remain $\alpha = \beta = 1$. Please note that the algorithm
currently only takes sentences into account which were found in a single article.
In the future, the MLP system will examine sets of closely related articles.
This will leverage the problem that distributional properties will be volatile on
term universes with very few members (e.g., term frequencies in a single sentence).

\paragraph{Implementation.}

We implemented the MLP processing system \cite{github} as Stratosphere
data-flow using Java which allows for scalable execution, application of complex
higher order functions, and easy integration of third party tools such as
Stanford NLP and the Mylyn framework for mark-up parsing.

\tikzset{actor/.style={
        rectangle,
        minimum size=6mm,
        very thick,
        draw=gray!50!black!50,
        top color=white,
        bottom color=gray!50!black!20
    },
    arrow/.style={
        -latex, thick, shorten <=2pt,shorten >=2pt
    }
}
\begin{figure}[H]
	\begin{tikzpicture}[node distance=5mm and 8mm]
		\node (Input) [align=center]{Wiki Dumps};
		\node (DocumentParser) [actor, right=of Input, align=center] {\emph{Map}\\\textbf{Parser}};
		\node (Candidates) [actor, right=of DocumentParser, align=center] {\emph{CoGroup}\\\textbf{Kernel}};
		\node (Sentence) [actor, above=of DocumentParser, align=center] {\emph{Map}\\\textbf{Tagger}};
		\node (Filter) [actor, right=of Candidates, align=center] {\emph{Reduce}\\\textbf{Filter}};
		\node (Output) [right=of Filter,align=center]{Raw\ Candidates};
		\draw[arrow] (Input)--(DocumentParser);
		\draw[arrow] (DocumentParser)--(Sentence);
		\draw[arrow] (DocumentParser)--(Candidates);
		\draw[arrow] (Sentence.east)--(Candidates.north);
		\draw[arrow] (Candidates)--(Filter);
		\draw[arrow] (Filter)--(Output);
	\end{tikzpicture}
\caption{Data flow of the Stratosphere program}
\end{figure}
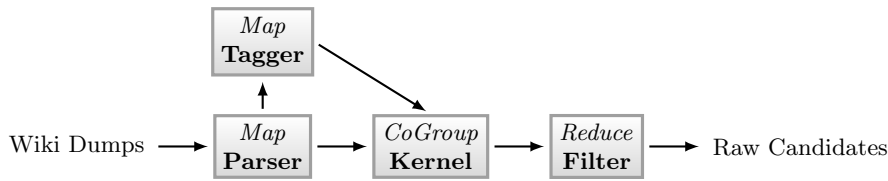

\section{Evaluation}

\subsection{Identifier Retrieval}
\label{ir}

Throughout our experiments, we made some observations that had an impact on the
accuracy of retrieving the correct set of identifiers. First of all, people
tend to use \emph{texvc} only as a typesetting language and neglect its semantic capabilities. For example,
\texttt{\textbackslash text\{log\}} is more often used than the correct
operator \texttt{\textbackslash log}. Another problem is that sometimes people
use indices as a form of `in field' annotation, like $T_{before}$ and
$T_{after}$. The identifier $T$ is defined in the surrounding text, but
neither $T_{before}$ nor $T_{after}$. There are more ambiguities. For example
the superscripted 2 in $x^{2}$ and $\sigma^{2}$ can be interpreted as the
power or as a part of the identifier. Another ambiguity is that the
multiplication sign can be omitted, so that it is undecidable for a naive
program whether $ab^{2}$ contains one or two identifiers.

We took a very conservative approach and preprocessed all formulas. The
\TeX~command \texttt{\textbackslash text\{\}} blocks along with subscriptions
containing more than a single character will be removed before analysis.
Superscripts will also be ignored in terms of being a part of the identifier.
Moreover, we created a comprehensive blacklist to improve the results further.
Identifier like `a', `A', and `I', which are also very common in the English
language, could be easily matched by our processor in the surrounding text,
and therefore, will also be blacklisted. Additionally, we blacklist common
mathematical operators, constants, and functions.

We took a sample of 30 random articles and counted all matches by hand. The
resulting estimates for the identifier retrieval performance are
\emph{Recall: 0.99} and \emph{Precision: 0.86}, which satisfy our information
needs, as we are mostly interested in recall at this stage.

\subsection{Description Retrieval}
We ran our program on a dataset of about 20,000 articles, all containing
\texttt{<math/>} tags, and retrieved about 550,000 candidate relations. The
most common definition relations are listed in table \ref{tcommon}.

\begin{table}[H]
	\begin{center}
	\begin{tabular}{| l | p{6.8cm} | l |}
		\hline
		Identifier & Descriptions & Count\\
		\hline
		$n$ & number & 1709 \\
		$t$ & time & 1480 \\
		$M$ & mass & 1042 \\
		$r$ & radius & 752 \\
		$T$ & temperature & 666 \\
		$\theta$ & angle & 639 \\
		$G$ & group & 635 \\
		\hline
	\end{tabular}
	\end{center}
\caption{\label{tcommon}Most common definition relations}
\end{table}

\paragraph{Observations.} We observed some poorly ranked relations. For
example, in the fragment \texttt{`where $\phi$( $r_{i}$ ) is the electrostatic
potential'}, the distance is 
$\Delta(\phi, \mathsf{electrostatic\:potential} ) = 6$. This is due
to counting brackets and function arguments as words. Also wrongly tagged
words like `Hamiltonian' as an adjective leads to false negatives.

\subsubsection{Comparative Evaluation}

At the start of our project there were no gold standard datasets available to
measure the performance of identifier definition extractors. Thus, we created
one on our own. This is a very time consuming job. At the moment, the dataset
only contains two large articles (revision ids are included) with around 100
identifier definitions. This dataset is also available on the project
repository.

As in many articles, those in the evaluation dataset contain identifiers
whose description cannot be retrieved. This is due to two reasons. First and
foremost, the identifier found in a formula is never mentioned in the
surrounding text, and therefore, no description can be extracted. Second, the
identifier is somehow ambiguous (see Section~\ref{ir}) and has been dropped. Most
notably, identifiers like $I_{xx}$ will be discarded because of an ambiguous
index that contains multiple letters.

Unfortunately 32 out of 99 identifiers from our dataset fall into that category.
We've decided to evaluate the performance of the remainder, as those 32 do not
convey any conceptual flaws. From the users standpoint, the overall performance
(in terms of recall) of such a system would be rather annoying. As we are only
interested in evaluating the performance of the \emph{MLP Ranking} algorithm itself,
it is safe to ignore those 32 identifiers.

\begin{table}[H]
\vspace{-5pt}
	\begin{center}
		\begin{tabular}{| l | l | l | l |}
			\hline
			 & MLP-Ranking ($k=1$) & MLP-Ranking ($k=2$) & Pattern Matching \\
			\hline
			Precision &  0.872  &  0.915  &  0.911  \\
			Recall    &  0.839  &  0.892  &  0.733  \\
			\hline
		\end{tabular}
	\end{center}
\caption{Evaluation results. Note: $k$ equals the amount of the top ranked candidate definitions.}
\vspace{-20pt}
\end{table}

Our results show that the unoptimized MLP approach keeps up with the
performance of the simple pattern matcher. Furthermore, we observed that it is
more robust in terms of recall, as it is less vulnerable to small changes in
the sentence structure.

\section{Further work}

Our original intuition was to discover grammatical patterns like
\texttt{`\emph{<identifier>} indicates/stands for/denotes
\emph{<description>}'} based on the statistical findings. However, our
impression is that this would not lead to significant performance gain.

The distance measure $R_{\sigma_d}$ fails for the example of
Fig.~\ref{fig:screenshot} since $\Delta(\mathsf{energy},E) =
\Delta(\mathsf{mass},m) = 2$. Unfortunately, one cannot simply detect
punctuation marks and introduce some kind of directed associativity
(e.g., inflicting a penalty on the ranking if the candidate relations spans over a
comma). This leads to whole classes of relation `types' (in terms of the
grammatical structure) never being retrieved. We plan to mitigate this problem
by taking more closely related scientific articles (based on their specific
fields) into consideration and count the frequencies of the candidate
relations. The intuition behind this is, that articles of the same scientific
field will likely use the same definition for the identifiers. Moreover,
we also hope to resolve the problem of `dangling' identifiers (those not
mentioned in the article itself), as they might be described in related
articles.

Currently, we use the ranking $R$ to identify the most probable description-
identifier tuple on each article, even if it occurs multiple times on the page.
For example, in the `Mass-energy equivalence' article, 21 sentences contain the
combination of the identifier $E$ and the noun `energy'. A promising approach,
is to use $R^\Sigma=\sum_{i=1}^n 2^{-i} R_i,$ where $R_i$ is a sorted list.
Here, $R_1$ is the highest ranked definition for that relation according to the
current measure $R$.  A systematic approach for determining a wise choice of the
ranking parameters should significantly improve the overall performance of our
system.

\section{Conclusion}

Our experiments showed that selecting candidates according to their POS tags
combined with numerical statistics about the text surface, can lead to quality
results. However, this approach is only applicable under certain conditions.
For identifiers which are seldom seen, our statistical approach tends to fail.
In that situation, other methods, especially supervised ones, are preferred.
Unfortunately, many of them require a labelled test corpus to measure the
performance of a classifier that could be trained with our generated data.
Currently, we are planning to use the NTCIR-Math-10 Task, Math Understanding
Subtask gold standard dataset \cite{overview} for a comparable evaluation.

During this project we had the impression that one could discover `namespaces'
(sets of documents, that use the same definitions for identifier) to aid in
the retrieval process. Robert Pagel is currently working on this topic
for his diploma thesis.

\paragraph*{Acknowledgments.}

Thanks to Howard Cohl for proofreading the paper and to Holmer Hemsen, the
course instructor of the database project course at TU-Berlin in Fall 2012.
The implementation and a first draft of this paper was completed in the
duration of this course.

\let\oldclearpage\clearpage
\let\clearpage\relax
\vspace{10 mm}
\bibliography{mlp-papers}
\let\clearpage\oldclearpage


\end{document}